\documentclass[aps,twocolumn,showpacs]{revtex4}
\usepackage{graphicx}
\usepackage{amssymb}
\usepackage{epstopdf}
\usepackage[usenames]{color}
\begin{document}

\title{Spout States in the Selective Withdrawal System}
\author{Sarah C. Case}
\affiliation{The James Franck Institute and Department of Physics, University of Chicago, Chicago, Illinois 60637}
\author{Sidney R. Nagel}
\affiliation{The James Franck Institute and Department of Physics, University of Chicago, Chicago, Illinois 60637}

\begin{abstract}

In selective withdrawal, fluid is withdrawn through a nozzle suspended above the flat interface separating two immiscible, density-separated fluids of viscosities $\nu_{upper}$ and $\nu_{lower} = \lambda \nu_{upper}$.  At low withdrawal rates, the interface gently deforms into a hump.  At a transition withdrawal rate, a spout of the lower fluid becomes entrained with the flow of the upper one into the nozzle.  When $\lambda = 0.005$, the spouts at the transition are very thin with features that are over an order of magnitude smaller than any observed in the humps.  When $\lambda = 20$, there is an intricate pattern of hysteresis and a spout appears which is qualitatively different from those seen at lower $\lambda$.  No corresponding qualitative difference is seen in the hump shapes.  


\end {abstract}

\maketitle

	Fluids change shape easily in response to stress.  In some cases, the change is so dramatic that the fluid interface changes topology.  An example is fluid entrainment by selective withdrawal.  Here, a change in topology occurs when shear stresses cause the initially smooth interface between two immiscible fluids to erupt into a spout that pierces through the entraining fluid.   If the shear is reduced, the spout collapses.  In this paper, we address the nature of this topological change by characterizing the spout shapes near the transition.

	 It is tempting to think of such topological changes as analogous to thermodynamic phase transitions.  Some fluid transformations, such as drop breakup, proceed as some physical dimension approaches zero, becoming much smaller than the macroscopic dimensions of the flows\cite{Zeff_2000, Eggers_1997, Lister_1998, Cohen_2001}.  Such a separation of length scales often leads to universal behavior as the dynamics approach a singularity where physical quantities diverge\cite{Constantin_1993, Goldstein_1993, Bertozzi_1996}.  This resembles second-order or critical behavior.  While this framework is appealing, it is imperfect.  It was recently discovered that ``critical" fluid transitions can exhibit a broader set of behaviors than their thermodynamic counterparts, and that not all fluid singularities obey universal dynamics\cite{Doshi_2003, Keim_2006}.  Other fluid transformations, such as selective withdrawal, proceed via a discontinuous jump and thus resemble first-order hysteretic transitions.  This paper explores the nature of this type of topological change. 

	   In our experiments, schematically shown in Fig. 1, a nozzle is suspended a height $S$ above the flat interface between two immiscible, density-separated fluids.  Fluid is withdrawn at a flow rate $Q$ through the nozzle, initially deforming the interface into an axisymmetric hump.  If $S$ is held fixed and $Q$ is increased, the hump will grow sharper until, at a transition flow rate, a spout of the lower fluid becomes entrained with the flow of the upper fluid into the nozzle.  There is hysteresis in this transition, and once the spout is formed, decreasing $Q$ past a second, lower transition flow rate causes the spout to collapse.	 

\begin{figure}
\centering
\includegraphics[width=80 mm]{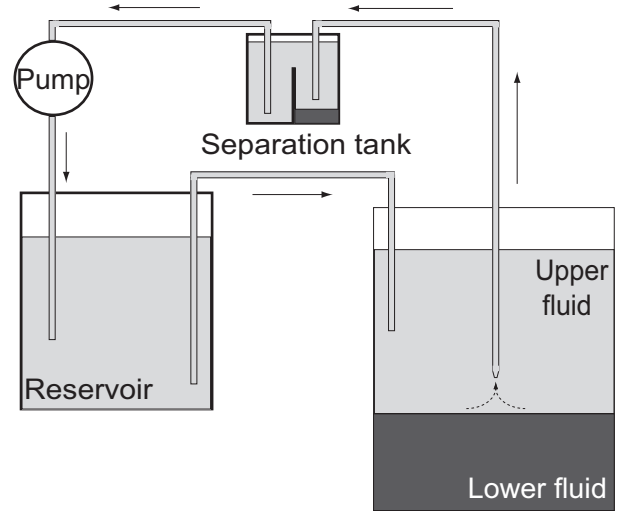}
\caption{Experimental setup.  Two immiscible fluids are density separated in a tank 20 cm x 20 cm x 30 cm.  Fluid is withdrawn through a nozzle a distance $S$ above the unperturbed interface using a  miniature gear pump.   Before reaching the pump, the lower fluid sediments out in an airtight separation tank, which also serves to damp pressure variations in the flows.  The upper fluid is deposited in a reservoir and siphoned back to the main tank, maintaining the upper fluid at a constant depth of 15-20 cm.  The apparatus is illuminated from the rear, and a CCD camera images the interface which is then traced using the programs ImageJ (NIH) and Pro Fit (Quansoft). }
\label{expt}
 \end{figure}
	   
	 We observe different behavior depending on whether the selective withdrawal transition is approached from low $Q$, in the hump state, or from high $Q$, in the spout state.  Previous experiments\nocite{Lister_1988}\cite{Cohen_2002, Cohen_2004} and simulations\cite{MKB_2005} have described the approach to this transition from the hump state, analyzing it either as a weakly first-order thermodynamic phase transition\cite{Cohen_2002, Cohen_2004} or as a saddle-node bifurcation\cite{MKB_2005}.   
	 	 
	 In this paper we analyze the approach to the transition from the spout.  We focus on two systems with different ratios $\lambda$ between the upper and lower fluid viscosities, $\nu_{upper}$ and  $\nu_{lower} = \lambda \nu_{upper}$, respectively.   When $\lambda = 0.005$, the spouts exhibit length scales near the transition that are an order of magnitude smaller than any observed in the humps.  These small scales allow the entrainment of extremely thin fluid threads, which are potentially useful in many applications, such as the coating of biological tissue for the purposes of transplantation.\cite{Cohen_2001_S, Wyman_2006}  On the other hand, when $\lambda = 20$, we observe a much thicker and qualitatively different spout.  By contrast this increase in the viscosity ratio does not cause a corresponding qualitative change in the hump state \cite{TBP}.  In addition, at this value of $\lambda$ two distinct types of spouts appear at the same flow rates but in different hysteretic regimes.
 
\begin{figure}
\centering
\includegraphics[width=80 mm]{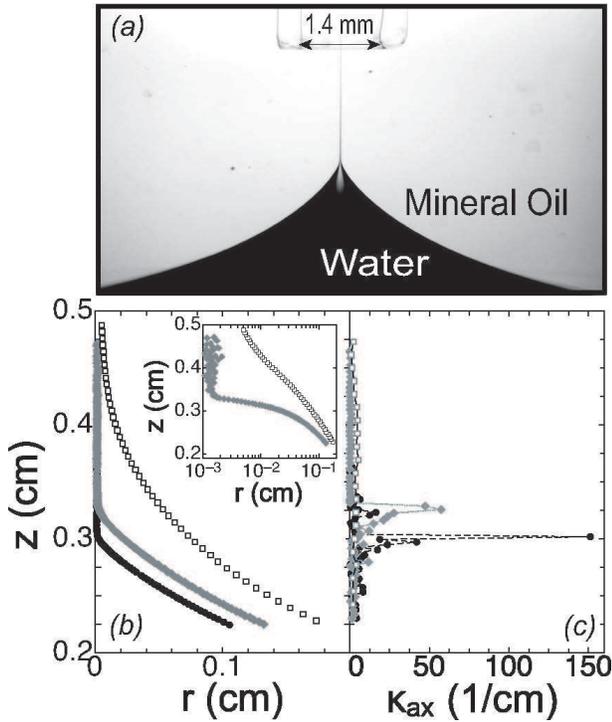}
\caption{Spout shapes at $\lambda = 0.005$.  The upper fluid is heavy mineral oil with $\nu_{upper} = 198$ cSt,  $\rho_{upper} = 0.87$ g/ml.  The lower fluid is deionized water with $\nu_{lower} = 1.000$ cSt,  $\rho_{lower} = 0.998$ g/ml.  The surface tension between the fluids is $\gamma = 35$ dyne/cm\cite{Cohen_2004}.   (a) Image of a typical spout.(b) $z$ versus $r$, shown for three spouts with $S = 0.51$ cm:  $(Q - Q^{*})/Q^{*} = 0.019$ (black circles), $(Q - Q^{*})/Q^{*} = 0.65$ (gray diamonds), $(Q - Q^{*})/Q^{*} = 2.34$ (open squares).  The outlines are averaged for every three pixels in $r$ and $z$. $S$ changes by less than $0.01 \%$ while acquiring each data point.  Inset shows $z$ versus $log(r)$.  (c) $z$ versus $\kappa_{ax}$ is shown for the same values of $S$ and $Q$ as in (b).} 
\end{figure} 

	We first examine the spout shapes near the transition when $\lambda = 0.005$.  A typical spout profile is seen in Fig. 2a.  Gravity forces the interface to be horizontal far from the nozzle, and the entraining flows force it to be vertical inside the nozzle.  We analyze the shape of the spatial region that connects these two asymptotic behaviors.  As $Q$ is varied, the shape of this connecting region changes, as shown in Fig. 2b.  At low $Q$, the spout resembles a thin thread attached to a broad base, with a localized region of high curvature connecting the two structures.  When $Q$ is increased, the spouts become smoother, approaching a nearly logarithmic shape as shown in the inset where $z$ is plotted versus $\log{r}$.
	
	We characterize these profiles by calculating the axial and azimuthal curvatures
\begin{equation}
 \kappa_{ax}(z) = \frac{\frac{d^{2}r}{dz^{2}}}{(1 + ( \frac{dr}{dz})^{2})^{3/2}}\ ; \qquad \kappa_{az}(z) = \frac{1}{r(z)}.
\end{equation}	
Very near the transition to the hump, the spout exhibits a sharp peak in $\kappa_{ax}$, as seen in Fig. 2c.  As $Q$ increases and the spouts become smoother, the peak in $\kappa_{ax}$ disappears below our resolution.  We characterize the evolution of the spout as the transition is approached by examining the following length scales: $z_{p}$ (the $z$ location of the peak in $\kappa_{ax}$),  $R_{p} \equiv 1/\kappa_{ax}(z_{p}) $, and $ r_{p} \equiv r(z_{p}) $.

\begin{figure}
\centering
\includegraphics[width=80 mm]{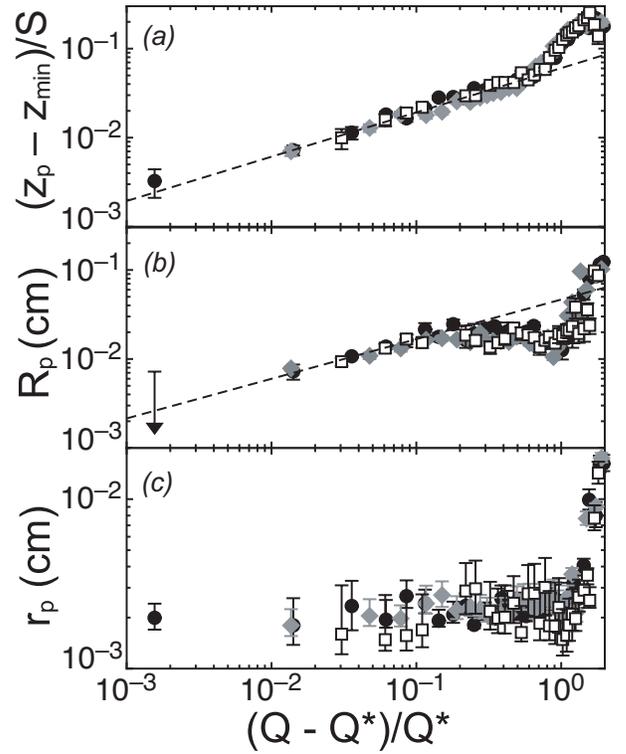}
\caption{Evolution of profile parameters as flow rate is decreased towards the transition.  (a) $\frac{z_{p} - z_{min}}{S}$ versus $(Q-Q^{*})/Q^{*}$.     (b) $R_{p}$ versus $(Q-Q^{*})/Q^{*}$.   (c) $r_{p}$ versus $(Q-Q^{*})/Q^{*}$.  In each case, three values of $S$ are shown: $S = 0.51$ cm (black circles), $S = 0.44$ cm (gray diamonds), and $S = 0.32$ cm (open squares).  Dashed lines are power law fits. } 
\end{figure} 
	
	In Fig. 3, we plot these profile parameters as functions of the flow rate $Q$ for three values of $S$.  Near the transition, $z_{p}$ approaches a minimum value $z_{min}$ as Q approaches a value $Q^{*}$.  Using $z_{min}$ and $Q^{*}$ as fitting parameters, we find that plotting $\log{(z_{p}- z_{min})/S}$ versus $\log{(Q-Q^{*})/Q^{*}}$ collapses the data onto a single curve, as shown in Fig. 3a, whereas the data for $R_{p}$ and $r_{p}$ collapse without scaling by $S$, as seen in Figs. 3b and 3c.  We search for power-law scaling in the same way as was done in the analysis of the approach to the transition from the hump state\cite{Cohen_2004}.  For ($(Q-Q^{*})/Q^{*} < 0.6$), $(z_{p} - z_{min})/S \propto ((Q- Q^{*})/Q^{*})^{0.50 \pm 0.08}$.  Using $Q^{*}$ determined from this fit, we find that $R_{p} \propto ((Q - Q^{*})/Q^{*})^{0.45 \pm 0.15}$.   In this case, our fitting range is smaller: $(Q-Q^{*})/Q^{*} < 0 .1$.  Although we are able to place an upper limit on the point closest to the transition, our imaging does not allow us to determine $R_{p}$ more precisely.  Finally, for $(Q - Q^{*})/Q^{*} < 1.5$, $r_{p}$ remains constant: $r_{p} = 22 \ \mu m \ \pm 4 \ \mu m$.  All three quantities show large departures from power-law behavior far from the transition.  
	
	Previous studies of the approach to the transition from low $Q$ found that the radius of curvature of the hump tip could be much smaller than the other lengths characterizing the system.  For Ê$\lambda = 0.005$, the largest separation of length scales $(h_{c}\kappa)_{hump}$ was $~10$(we estimate $h_{c}$ from the largest $h_{max}$ value shown)\cite{Cohen_2004} .   Applying a similar analysis to the approach from high $Q$, we find that $z_{min}/R_{p}$ is $~40$.  We note, however, that $R_{p}$ is not the smallest length scale observed for this system.  The azimuthal curvature $r_{p}$ is significantly smaller than $R_{p}$, so that $z_{min}/r_{p}$ is greater than $150$.
	
	When the viscosity ratio is increased to $\lambda = 20$, we observe a qualitatively different spout.   These spouts are very thick, and the flows penetrate deeply into the bulk of the lower fluid. This is shown in Fig. 4a, where the flows, obtained by dyeline tracing inside the lower fluid, are superimposed on a typical spout image.  These flows are distinct from those seen when $\lambda = 0.005$, where the flows are primarily along the interface\cite{Wyman_2006}.  We therefore refer to the broad spouts as ``bulk-flow" spouts.  We note that the bulk-flow spout can be so thick that it fills the entire orifice when the nozzle diameter, $D$, is small.  We restrict our analysis to $D$ large enough that this does not occur.   

\begin{figure}
\centering
\includegraphics[width=80 mm]{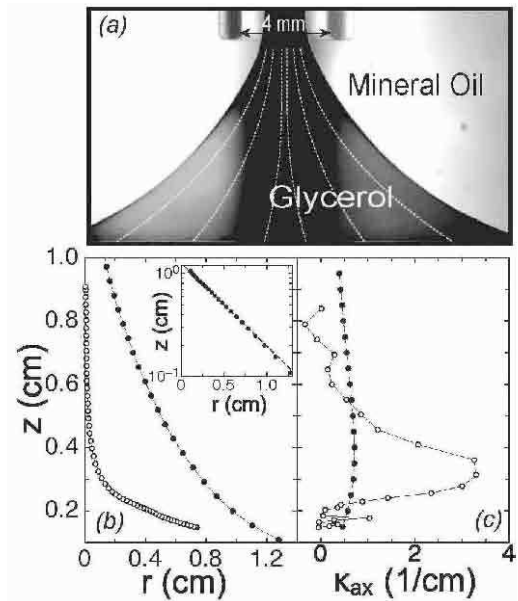}
\caption{Spout shapes at $\lambda = 20$.  The upper fluid is light mineral oil with $\nu_{upper} = 48$ cSt,  $\rho_{upper} = 0.85$ g/ml.  The lower fluid is glycerol with $\nu_{lower} = 980$ cSt,  $\rho_{lower} = 1.26$ g/ml.  The surface tension between the two fluids was measured to be $\gamma = 27 \pm 1$ dyne/cm using the pendant drop method.  (a) image of a typical bulk-flow spout.  Flow lines were traced from dyelines in another image at the same $S$ and similar $Q$ and superimposed on the image shown.  (b) $z$ versus $r$ for a bulk-flow spout (filled circles) at $Q = 57.1$ ml/s and $S = 1.05$ cm. and a hysteretic surface-flow spout (open squares) at the same value of $S$ but with $Q = 54.7$ ml/s.  Inset shows exponential fit to the bulk-flow spout, $z=1.26e^{-r/0.54}$. (c) $z$ versus $\kappa^{ax}$ for same data shown in (b).}
\end{figure} 

	At $\lambda = 20$, a spout closely resembling those seen at small $\lambda$ also exists in a narrow hysteretic region at the same values of $S$ and $Q$ as does the bulk-flow spout.  We are able to observe these spouts by carefully manipulating the flow rate.  At the transition flow rate, an unstable thread of the lower fluid becomes entrained in the upper one.  After a time lag, which can be as long as several seconds, this initial spout begins to widen, culminating in the steady bulk-flow spout.  However, if $Q$ is rapidly reduced before the initial spout widens significantly, a stable spout as much as two orders of magnitude thinner than the bulk-flow one is produced.  This thin spout exhibits flow patterns in the lower fluid similar to those seen at $\lambda = 0.005$.	
	
	The profiles of these two spouts are distinctly different, as seen in Fig. 4b.  In the inset to Fig. 4b, the bulk-flow profiles are fit to $z \propto e^{-r/r_{0}}$.   The data deviates very slightly from this form very near the straw entry as well as near the horizontal interface, but is in excellent agreement in the intermediate region. This fit is used to calculate $\kappa_{ax}(z)$ for these spouts. The decay length, $r_{0}$, changes only slightly as a function of $D$:  for $D = 0.40$ cm, $r_{0} = 0.54$ cm $\pm$ $0.02$ cm, while for $D = 0.80$, $r_{0}= 0.44$ cm $\pm$ $0.02$ cm.   In contrast to this exponential, the hysteretic surface-flow spout profiles are closer to logarithmic, and are similar to those seen in the inset to Fig 2b for $\lambda = 0.005$. The hysteretic surface-flow spout also displays a sharper peak in $\kappa_{ax}$ than the bulk-flow spout, as shown in Fig. 4c.
	
	A clearly defined transition exists between the two spouts. At a threshold flow rate, the stable surface-flow spout becomes unstable and rapidly widens into a bulk-flow spout with no stable intermediate structures. Thus, there are transitions between three distinct steady states.  In Fig. 5, we show these transitions, including the relevant hystereses, on a phase diagram of the transition flow rate $Q_{t}$ versus $S$.  As seen in Fig. 5a, the transition from hump to bulk-flow spout can be fit to $S = (0.29 \pm 0.01)(Q_{t})^{0.32 \pm 0.03}$.  The hysteresis in the hump to bulk-flow spout transition is large, especially at low $S$, and is observed to vary with $D$.   Fig. 5b shows a narrow but well-defined region in which, depending on the initial state of the system, surface-flow spouts, bulk-flow spouts, and humps can all be observed.  This region drops below experimental resolution at $S$ = 35 ml/s.

\begin{figure}
\centering
\includegraphics[width=80 mm]{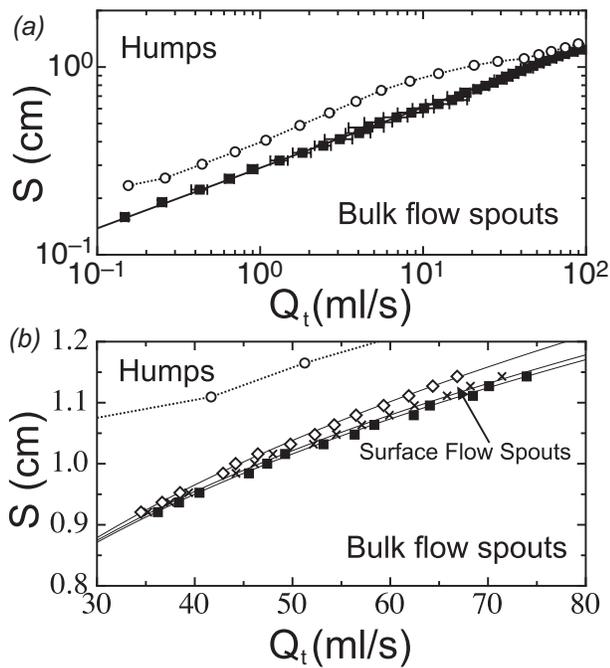}
\caption{Phase diagram for $\lambda = 20$ showing the different hysteresis regions.   (a) The transitions from hump to bulk-flow spout (filled squares) and from bulk-flow spout to hump (open circles) are shown.  The solid line is $S = (0.29) Q_{t}^{0.32}$.  (b)  A magnified view of the phase diagram at high $Q_{t} $ showing, for a representative data set, the surface-flow spout to hump transition (open diamonds) and the surface-flow spout to bulk-flow spout transition (x's).  The lines are guides to the eye.}  
\end{figure} 

In conclusion, the independent variation of the viscosity ratio $\lambda$ and the flow rate $Q$ dramatically changes the spout profiles.  At $\lambda = 20$ there is a broad bulk-flow spout which disappears for smaller $\lambda$.  Thin surface-flow spouts, seen at low $\lambda$, are only visible in a limited hysteretic region at $\lambda = 20$.  Cohen\cite{Cohen_2004} observed no qualitative difference in the hump shapes near the transition for $\lambda$ between $10^{-3}$ and $1.7$.  Similar analysis of our data extends the range to $\lambda = 20$\cite{TBP}.  Thus, a qualitative change in the spout, not seen in the hump, is found by varying $\lambda$.   Decreasing the other control parameter, $Q$, while fixing $\lambda=0.005$, causes the spout to narrow drastically as the transition is approached.  This produces a much greater separation of length scales in the spout than found in the hump.

Unlike the hump interface, the spout interface is not bounded in the vertical direction.  Thus, the spout has two asymptotic regimes which must be matched to each other by the dynamics: at large radius it is constrained by gravity to be horizontal and at large heights it is constrained by the flows in the nozzle to be vertical.  The increase in the separation of length scales as the transition is approached can be viewed as a decrease in the spatial width connecting these constraints.  At high $Q$, the axial curvature, $\kappa_{ax}$, is nearly constant in $z$.  As $Q$ is decreased to the point where the spout disappears, the transition region becomes increasingly localized, leading to a sharp peak in $\kappa_{ax}$.  However, the spatial transition region never collapses to zero and $\kappa_{ax}$ is always cut off at a finite value.  The degree to which the selective-withdrawal transition can approach a continuous one with a singularity is limited by the minimum extent to which the zone matching the two constraints can shrink to zero.
	   	 	 
	We are grateful to W. W. Zhang, F. Blanchette, M. Kleine-Berkenbusch, J. L. Wyman, and K. Walker for helpful discussions.  This research was supported by NSF MRSEC DMR-0213745 and NSF DMR-0352777.


\begin{thebibliography}{16}
\expandafter\ifx\csname natexlab\endcsname\relax\def\natexlab#1{#1}\fi
\expandafter\ifx\csname bibnamefont\endcsname\relax
  \def\bibnamefont#1{#1}\fi
\expandafter\ifx\csname bibfnamefont\endcsname\relax
  \def\bibfnamefont#1{#1}\fi
\expandafter\ifx\csname citenamefont\endcsname\relax
  \def\citenamefont#1{#1}\fi
\expandafter\ifx\csname url\endcsname\relax
  \def\url#1{\texttt{#1}}\fi
\expandafter\ifx\csname urlprefix\endcsname\relax\def\urlprefix{URL }\fi
\providecommand{\bibinfo}[2]{#2}
\providecommand{\eprint}[2][]{\url{#2}}

\bibitem[{\citenamefont{Zeff et~al.}(1994)\citenamefont{Zeff, Kleber, Fineberg,
  and Lathrop}}]{Zeff_2000}
\bibinfo{author}{\bibfnamefont{B.~W.} \bibnamefont{Zeff}},
  \bibinfo{author}{\bibfnamefont{B.}~\bibnamefont{Kleber}},
  \bibinfo{author}{\bibfnamefont{J.}~\bibnamefont{Fineberg}}, \bibnamefont{and}
  \bibinfo{author}{\bibfnamefont{D.~P.} \bibnamefont{Lathrop}},
  \bibinfo{journal}{Nature (London)} \textbf{\bibinfo{volume}{403}},
  \bibinfo{pages}{401} (\bibinfo{year}{1994}).

\bibitem[{\citenamefont{Eggers}(1997)}]{Eggers_1997}
\bibinfo{author}{\bibfnamefont{J.}~\bibnamefont{Eggers}},
  \bibinfo{journal}{Rev. Mod. Phys} \textbf{\bibinfo{volume}{69}},
  \bibinfo{pages}{865} (\bibinfo{year}{1997}).

\bibitem[{\citenamefont{Lister and Stone}(1998)}]{Lister_1998}
\bibinfo{author}{\bibfnamefont{J.~R.} \bibnamefont{Lister}} \bibnamefont{and}
  \bibinfo{author}{\bibfnamefont{H.~A.} \bibnamefont{Stone}},
  \bibinfo{journal}{Phys. Fluids} \textbf{\bibinfo{volume}{10}},
  \bibinfo{pages}{2758} (\bibinfo{year}{1998}).

\bibitem[{\citenamefont{Cohen and Nagel}(2001)}]{Cohen_2001}
\bibinfo{author}{\bibfnamefont{I.}~\bibnamefont{Cohen}} \bibnamefont{and}
  \bibinfo{author}{\bibfnamefont{S.~R.} \bibnamefont{Nagel}},
  \bibinfo{journal}{Phys. Fluids} \textbf{\bibinfo{volume}{2001}},
  \bibinfo{pages}{3533} (\bibinfo{year}{2001}).

\bibitem[{\citenamefont{Constantin et~al.}(1993)\citenamefont{Constantin,
  Dupont, Goldstein, Kadanoff, Shelley, and Zhou}}]{Constantin_1993}
\bibinfo{author}{\bibfnamefont{P.}~\bibnamefont{Constantin}},
  \bibinfo{author}{\bibfnamefont{T.~F.} \bibnamefont{Dupont}},
  \bibinfo{author}{\bibfnamefont{R.~E.} \bibnamefont{Goldstein}},
  \bibinfo{author}{\bibfnamefont{L.~P.} \bibnamefont{Kadanoff}},
  \bibinfo{author}{\bibfnamefont{M.~J.} \bibnamefont{Shelley}},
  \bibnamefont{and} \bibinfo{author}{\bibfnamefont{S.~M.} \bibnamefont{Zhou}},
  \bibinfo{journal}{Phys. Rev. E} \textbf{\bibinfo{volume}{47}},
  \bibinfo{pages}{4169} (\bibinfo{year}{1993}).

\bibitem[{\citenamefont{Goldstein et~al.}(1993)\citenamefont{Goldstein, Pesci,
  and Shelley}}]{Goldstein_1993}
\bibinfo{author}{\bibfnamefont{R.~E.} \bibnamefont{Goldstein}},
  \bibinfo{author}{\bibfnamefont{A.~I.} \bibnamefont{Pesci}}, \bibnamefont{and}
  \bibinfo{author}{\bibfnamefont{M.~J.} \bibnamefont{Shelley}},
  \bibinfo{journal}{Phys. Rev. Lett.} \textbf{\bibinfo{volume}{70}},
  \bibinfo{pages}{3043} (\bibinfo{year}{1993}).

\bibitem[{\citenamefont{Bertozzi et~al.}(1996)\citenamefont{Bertozzi, Brenner,
  Dupont, and Kadanoff}}]{Bertozzi_1996}
\bibinfo{author}{\bibfnamefont{A.~L.} \bibnamefont{Bertozzi}},
  \bibinfo{author}{\bibfnamefont{M.~P.} \bibnamefont{Brenner}},
  \bibinfo{author}{\bibfnamefont{T.~F.} \bibnamefont{Dupont}},
  \bibnamefont{and} \bibinfo{author}{\bibfnamefont{L.~P.}
  \bibnamefont{Kadanoff}}, in \emph{\bibinfo{booktitle}{Trends and Perspectives
  in Applied Mathematics}}, edited by
  \bibinfo{editor}{\bibfnamefont{L.}~\bibnamefont{Sirovich}}
  (\bibinfo{publisher}{Springer (New York)}, \bibinfo{year}{1996}).

\bibitem[{\citenamefont{Doshi et~al.}(2003)\citenamefont{Doshi, Cohen, Zhang,
  Siegel, Howell, Basaran, and Nagel}}]{Doshi_2003}
\bibinfo{author}{\bibfnamefont{P.}~\bibnamefont{Doshi}},
  \bibinfo{author}{\bibfnamefont{I.}~\bibnamefont{Cohen}},
  \bibinfo{author}{\bibfnamefont{W.~W.} \bibnamefont{Zhang}},
  \bibinfo{author}{\bibfnamefont{M.}~\bibnamefont{Siegel}},
  \bibinfo{author}{\bibfnamefont{P.}~\bibnamefont{Howell}},
  \bibinfo{author}{\bibfnamefont{O.}~\bibnamefont{Basaran}}, \bibnamefont{and}
  \bibinfo{author}{\bibfnamefont{S.~R.} \bibnamefont{Nagel}},
  \bibinfo{journal}{Science} \textbf{\bibinfo{volume}{302}},
  \bibinfo{pages}{1185} (\bibinfo{year}{2003}).

\bibitem[{\citenamefont{Keim et~al.}(2006)\citenamefont{Keim, Zhang, and
  Nagel}}]{Keim_2006}
\bibinfo{author}{\bibfnamefont{N.}~\bibnamefont{Keim}},
  \bibinfo{author}{\bibfnamefont{W.~W.} \bibnamefont{Zhang}}, \bibnamefont{and}
  \bibinfo{author}{\bibfnamefont{S.~R.} \bibnamefont{Nagel}}
  (\bibinfo{year}{2006}), \bibinfo{note}{to be published}.

\bibitem[{\citenamefont{Lister}(1988)}]{Lister_1988}
\bibinfo{author}{\bibfnamefont{J.~R.} \bibnamefont{Lister}},
  \bibinfo{journal}{J. Fluid Mech.} \textbf{\bibinfo{volume}{198}},
  \bibinfo{pages}{231} (\bibinfo{year}{1988}).

\bibitem[{\citenamefont{Cohen and Nagel}(2002)}]{Cohen_2002}
\bibinfo{author}{\bibfnamefont{I.}~\bibnamefont{Cohen}} \bibnamefont{and}
  \bibinfo{author}{\bibfnamefont{S.~R.} \bibnamefont{Nagel}},
  \bibinfo{journal}{Phys. Rev. Lett.} \textbf{\bibinfo{volume}{88}},
  \bibinfo{pages}{074501} (\bibinfo{year}{2002}).

\bibitem[{\citenamefont{Cohen}(2004)}]{Cohen_2004}
\bibinfo{author}{\bibfnamefont{I.}~\bibnamefont{Cohen}},
  \bibinfo{journal}{Phys. Rev. E} \textbf{\bibinfo{volume}{70}},
  \bibinfo{pages}{026302} (\bibinfo{year}{2004}).

\bibitem[{\citenamefont{Kleine-Berkenbusch
  et~al.}(2006)\citenamefont{Kleine-Berkenbusch, Cohen, and Zhang}}]{MKB_2005}
\bibinfo{author}{\bibfnamefont{M.}~\bibnamefont{Kleine-Berkenbusch}},
  \bibinfo{author}{\bibfnamefont{I.}~\bibnamefont{Cohen}}, \bibnamefont{and}
  \bibinfo{author}{\bibfnamefont{W.~W.} \bibnamefont{Zhang}}
  (\bibinfo{year}{2006}), \bibinfo{note}{submitted to J. Fluid Mech.}

\bibitem[{\citenamefont{Cohen et~al.}(2001)\citenamefont{Cohen, Li, Hougland,
  Mrksich, and Nagel}}]{Cohen_2001_S}
\bibinfo{author}{\bibfnamefont{I.}~\bibnamefont{Cohen}},
  \bibinfo{author}{\bibfnamefont{H.}~\bibnamefont{Li}},
  \bibinfo{author}{\bibfnamefont{J.~L.} \bibnamefont{Hougland}},
  \bibinfo{author}{\bibfnamefont{M.}~\bibnamefont{Mrksich}}, \bibnamefont{and}
  \bibinfo{author}{\bibfnamefont{S.~R.} \bibnamefont{Nagel}},
  \bibinfo{journal}{Science} \textbf{\bibinfo{volume}{292}},
  \bibinfo{pages}{265} (\bibinfo{year}{2001}).

\bibitem[{\citenamefont{Wyman et~al.}(2006)\citenamefont{Wyman, Kizilel,
  Mrksich, Garfinkel, and Nagel}}]{Wyman_2006}
\bibinfo{author}{\bibfnamefont{J.~L.} \bibnamefont{Wyman}},
  \bibinfo{author}{\bibfnamefont{S.}~\bibnamefont{Kizilel}},
  \bibinfo{author}{\bibfnamefont{M.}~\bibnamefont{Mrksich}},
  \bibinfo{author}{\bibfnamefont{M.}~\bibnamefont{Garfinkel}},
  \bibnamefont{and} \bibinfo{author}{\bibfnamefont{S.~R.} \bibnamefont{Nagel}}
  (\bibinfo{year}{2006}), \bibinfo{note}{submitted to Small}.

\bibitem[{\citenamefont{Case and Nagel}(2006)}]{TBP}
\bibinfo{author}{\bibfnamefont{S.~C.} \bibnamefont{Case}} \bibnamefont{and}
  \bibinfo{author}{\bibfnamefont{S.~R.} \bibnamefont{Nagel}}
  (\bibinfo{year}{2006}), \bibinfo{note}{to be published}.

\end{thebibliography}
\end{document}